\documentclass[journal,onecolumn,draftcls,12pt]{IEEEtran}
\usepackage{amsmath,amssymb,amsfonts,amsthm,mathtools}
\usepackage{subfigure}

\ifCLASSINFOpdf
\else
\fi
\newcommand{\G}{\mathcal{G}}

\newcommand{\E}{\mathcal{E}}

\newcommand{\SA}{\mathcal{S}}

\usepackage{latexsym}

\begin{document}

\title{Opinion shaping in social networks using reinforcement learning}

\author{\IEEEauthorblockN{Vivek S.\ Borkar,\footnote{Work supported in part by a J.\ C.\ Bose Fellowship from the Government of India and a grant `Monte Carlo and Learning Schemes for Network Analytics' from Indo-French Centre for Promotion of Advanced Research (CEFIPRA).}}
\IEEEauthorblockA{Department of Electrical  Engineering\\
Indian Institute of Technology\\
Powai, Mumbai 400076, India\\
Email: borkar.vs@gmail.com\\}
\and
\IEEEauthorblockN{Alexandre Reiffers-Masson,}
\IEEEauthorblockA{Robert Bosch Centre for Cyberphysical Systems,\\
Indian Institute of Science\\
Bengaluru 560012, India\\
Email: reiffers.alexandre@gmail.com }
}

\maketitle

\begin{abstract}

Recent studies have demonstrated that the decisions of agents in society are shaped by their own intrinsic motivation, and also by the compliance with the social norm. In other words, the decision of acting in a particular manner will be affected by the opinion of society. This social comparison mechanism can lead to imitation behavior, where an agent will try to mimic the behavior of her neighbors. Using this observation, new policies have been designed, e.g., in the context of energy efficiency and transportation choice, to leverage social networks in order to improve altruism and prosocial behavior.
One policy is to use targeting strategies. Indeed, by changing the behavior of influential actors in a social network, it is possible to reshape the global behavior of agents towards more prosocial behavior. However, discovering who are the influential agents requires a lot of information, such as the matrix of interactions between agents.
In this paper, we study how to shape opinions in social networks when the matrix of interactions is unknown.
We consider classical opinion dynamics with some stubborn agents and the possibility of continuously influencing the opinions of a few selected agents, albeit under resource constraints. We map the opinion dynamics to a value iteration scheme for policy evaluation for a specific stochastic shortest path problem. This leads to a representation of the opinion vector as an approximate value function for a stochastic shortest path problem with some non-classical constraints.
We suggest two possible ways of influencing agents. One leads to a convex optimization problem and the other to a non-convex one. Firstly, for both problems, we propose two different online two-time scale reinforcement learning schemes that converge to the optimal solution of each problem. Secondly, we suggest stochastic gradient descent schemes and compare these classes of algorithms with the two-time scale reinforcement learning schemes. Thirdly, we also derive another algorithm designed to tackle the curse of dimensionality one faces when all agents are observed. Numerical studies are provided to illustrate the convergence and efficiency of our algorithms.

\end{abstract}

\IEEEpeerreviewmaketitle



\section{Introduction}

In recent times there has been increasing interest in non-price based mechanisms to improve society's behavior in the context of, e.g.,  energy efficiency or traffic behavior. These policies are usually less expensive to implement and can be politically feasible as opposed to price based policies. One example is the use of lottery with the distribution of coupons for energy efficiency  \cite{xia2017energycoupon} or for promoting off-peak usage of cars \cite{merugu2009incentive}. Another example is of leveraging social network for enhancing pro-social behavior. Indeed, social interactions can  impact  day to day decisions of an agent.
For instance, in transportation choice, several works (\cite{dugundji2005discrete,walker2011correcting}) have demonstrated that the preferences of people in the decision maker's peer group will impact her choice of mode of transport to work (public transport, bicycle, car). Another practical application concerns how to use social comparison to enhance energy efficiency \cite{schultz2018constructive}. One specific way of leveraging social network for such purposes is to exploit the word-of-mouth/imitation process in a network by using a targeted advertising campaign. Targeted advertising on a social network amounts to finding which agents should be convinced in a social network to be pro-social  in such a way that by imitation, the maximum number of agents in the whole social network will also be pro-social. Designing such a  targeting strategy, however, can be challenging because of: (1) computational issues, (2) unknown social network, (3) size or lack of convexity of the resulting optimization problem, and so on. The goal of this paper is to propose different algorithms based on reinforcement learning that address these issues.

Our initial model can be described as follows:  The society is composed of a finite set of agents and each agent has an opinion concerning a given pro-social action that she has to take. For instance, it could be the opinion concerning whether or not she should take the bus to work or how much she cares about energy efficiency of her apartment. Whether an agent performs or not the pro-social action,  the rest of the society that is "close" to her in the social network will observe her and vice versa. Therefore each agent will have a tendency to imitate her neighbors and vice versa.
A planner (government, owner of the social network) is interested in choosing which agents she will influence to shape the opinion in a given direction. The planner can influence an agent through two controls which will be described later on. The society is divided into three types of agents. The first set is composed of \textit{'stubborn' agents}. In this set, we assume that they have a given opinion and they will not be impacted by the social network. The second set of agents is composed of \textit{'uncontrolled' agents}. In this set, agents are influenced by the social network and their opinion will be impacted by the opinion of the others. However, the planner cannot directly influence them. The last set of agents is called the set of \textit{`controlled' agents}. This set is composed of agents that care about the opinion of their neighbors and also can be influenced by the planner. The goal of the planner is to shape the opinion of the social network by targeting specific agents from the latter group. We call this problem the \textit{opinion shaping problem}. One of the major drawbacks of this initial model is that for shaping the opinion of the society, the planner needs to know the influence matrix. Worse, even if the influence matrix is known, the number of agents is so big that it is not feasible to decide optimally which agent should be chosen. Finally, depending on how the planner can influence the users, the convexity of the problem can be lost.

Our initial model is inspired by \cite{Bimp,Consent}. In these papers, the influence matrix is assumed to be known. Moreover the authors do not consider efficient and decentralized  algorithms to solve the opinion shaping problem. In this paper, we extend these works further in order to address these issues. Throughout our paper, the main assumption is that the planner does not know the influence matrix but he observes when the agents interact. Using these observations, we are able to derive three reinforcement learning based algorithms that will address the aforementioned issues. The proposed algorithms are  decentralized. We also provide supporting simulations for the different settings. From a mathematical point of view, we provide the equivalence of the opinion-shaping problem with a stochastic shortest path problem and use this correspondence to motivate our algorithms and  their convergence. Additionally, we discuss several possible extensions and future directions.

\subsection{Organization and Main Results}

The remainder of this paper is organized into eight sections. In section \ref{sec: related works}, we discuss related works. Section \ref{sec: opinion shaping problem} introduces the opinion-shaping problem after first describing the model for the opinion adoption process. Section \ref{sec: first algo} is the main section of this paper. We prove the equivalence of the opinion adoption process with a stochastic shortest path problem. Using this equivalence, we propose a decentralized algorithm accounting for the fact that the influence matrix is unknown. This is followed by two other variants. The first one is using a unbiased estimator for the gradient and the second one is designed  to tackle the curse of dimensionality one faces when all agents are observed. In section \ref{sec: Convergence analysis}, we prove the convergence of our algorithms. In section \ref{sec: a more general model}, a new algorithm is proposed when the opinion-shaping problem is not convex. Numerical studies are discussed in \ref{sec: numerical experiments}.
Moreover, we compare the efficiency of the second algorithm, where all the agents are not observed, with the one where all the agents are observed. Finally, we study the efficiency of the annealing scheme for the non-convex opinion-shaping problem.  Section \ref{sec: further directions} concludes with pointers to some possible extensions of this work.

\section{Related works}\label{sec: related works}
The spread of opinions in social networks broadly falls into three families of mathematical models and the influence maximization  problem has been studied  in the context of each of these. The first category is cascade threshold models. In the last decade, initial models in this framework for control of user activity  were dedicated to maximization of influence \cite{guille2013information}, including the seminal work of Kempe et al. \cite{kempe2005influential}. One of the drawbacks of this line of work is that the state of each user is assumed to be finite,  often even binary. Another key limitation  is that it only focuses on the maximization of influence, which reduces its possible scope for  applications. Indeed, one may also be interested in other objectives such as minimum activity in a social network or diverse activity and not just activity maximization. Based on these, a second category of models has been proposed, e.g., by  Zha et al. in \cite{farajtabar2014shaping}, who define a new mathematical problem  dubbed the activity shaping problem. First, they use Hawkes processes to model the activity of users in a social network. Undeniably, in recent literature these point processes have proved to be a very effective method to capture users' activity \cite{zhou2013learning}. Secondly,  the authors consider an activity shaping problem wherein by controlling the exogenous rate vectors of the Hawkes processes, a central controller tries to minimize a convex function which depends on the expected overall instantaneous intensity of the processes. Since then, other extensions have been suggested, e.g., \cite{zarezade2017cheshire,wang2016stochastic}.

Our model of opinion propagation falls in the last category, viz., consensus models. For several years, much effort has been devoted to the study of users' activities in a social network within this framework. The seminal work of Degroot \cite{degroot1974reaching} proposes a simple model to capture the diffusion of opinion. In his paper, the author assumes that each agent at each instant will compute the average opinion of her neighbors including possibly herself, and then replace her current opinion by this average. Several extensions of the Degroot model have been considered recently, especially the opinion shaping part \cite{Bimp,Consent,masson2017posting}. In \cite{Bimp}, the authors assume that a planner can directly contaminate a user in the social network by sending him some messages. The user reads the messages according to a certain probability and with the remaining probability, she will sample from  the messages sent by her friends. In \cite{Consent}, the authors consider a more drastic control where they can freeze the opinion of a given user. Finally, in \cite{masson2017posting}, the authors suggest a control based on the reduction of the interaction between different agents of the social network.

Concerning the opinion shaping problem without the knowledge of the network, the authors in \cite{lin2015learning} propose a data-driven model and a learning algorithm in case of a cascade with a linear threshold model. In \cite{yadav2016using}, the authors suggest a Partially Observed Markov Decision Process (POMDP) framework in order to tackle the uncertainty over the topology of the network, again for the case of a cascade with linear threshold model.

To the best of our knowledge, our paper is the first one that tries to shape the opinion under resource constraints using a reinforcement learning approach, under the assumption that the opinion propagation is captured by a consensus model, but without the full knowledge of the topology.

\section{Preliminaries and model}\label{sec: opinion shaping problem}
We consider a social network given by a connected directed graph $\G = (\SA, \E)$ where $\SA$ is the set of its agents and $\E$ the set of edges. To each edge $(i,j) \in \E$ we assign a probability weight $p_{ij} > 0$ with $\sum_{\{j : (i,j) \in \E\}}p_{ij} = 1$. We set $p_{ij} = 0$ if $(i,j) \notin \E$. The total number of agents is equal to $I$.

With each agent $i \in \SA$, we also associate a process of valuations $x_i(k) \in [0, 1], n \geq 0$. Let $x(k)=[x_1(k),\ldots,x_I(k)]$ be the associated vector for each $k$. We write $\SA$ as a disjoint union $\SA = S\cup S_0\cup S_1$ of three sets. Each time instant $k$, an agent $i$ (more generally, a set of agents) from $\SA$ is selected and updates her valuation. The update mechanism will change depending on which set an agent belongs to:

 \textbf{Set of `stubborn' agents ($S_0$): } the agents that belong to this set have their valuation frozen at some fixed value for good, i.e., $i \in S_0 \Longrightarrow \forall k,\;x_i(k) \equiv h(i) \in [0, 1]$.

\textbf{Set of `uncontrolled' agents ($S_1$): } This set stands for agents for which their valuations evolves according to a gossip mechanism, but they are not amenable to external influence. In this case the valuation update of an agent $i\in S_1$ is captured by the following mechanism: agent $i$ polls a neighbor $\ell\in\SA$ with probability $p_{i\ell}$ and updates $x_i(k)$ to $x_i(k+1) = x_\ell(k)$.

\textbf{Set of `controlled' agents ($S$): } The remaining agents that constitute the set $S$ also evolve according to a gossip mechanism, but are amenable to external influence or `control'. With probability $\alpha_i\in(0,1)$, agent $i$ will be influenced directly by the planner and with probability $1-\alpha_i$, agent $i$ will be influenced by her peer group. When influenced by the planner, agent $i$ updates $x_i(k)$ to $x_i(k+1) = w_i(u_i)$, with $u_i \in \mathbb{R}_+$ without loss of generality and $w_i : \mathbb{R}_+ \mapsto [0, 1]$ are concave increasing and continuously differentiable maps. (Concavity captures the `diminishing returns' effect.)  If agent $i$ is influenced by her peer group, she polls a neighbor $\ell$ with probability $p_{i\ell}$ and updates $x_i(k+1)=x_\ell(k)$. 

To summarize, the overall dynamics is then described as follows. Suppose agent $i \in \SA$ performs an update at time $k$. Then
\begin{eqnarray*}
x_i(k+1) &=& \left\{\begin{array}{lll}w_i(u_i)&w.p.&\alpha_i\\
x_\ell(k)&w.p.&1-\alpha_i
\end{array}\right. \ i \in S, \\
x_i(k+1) &=& x_\ell(k), \ i \in S_1,\\
x_i(k+1) &=&h(i), \ i \in S_0,\\
x_j(k+1) &=& x_j(k) \ \forall \ j \neq i.
\end{eqnarray*}
Analogous scheme holds when more than one agent updates.

For each $i\in\SA$, when $k \to \infty$, $x_i^*=\lim_{k\rightarrow+\infty}E\left[x_i(k)\right]$ is the solution of the following fixed-point equation:
\begin{eqnarray*}
x_i^* &=& \alpha_iw_i(u_i) +   (1 - \alpha_i)\sum_{\ell\in\SA}p_{i\ell}x_\ell^*, \ i \in S,\\
x_i^* &=& \sum_{\ell\in\SA}p_{i\ell}x_\ell^*, \ i \in S_1,\\
x_i^*  &=&h(i), \ i \in S_0.
\end{eqnarray*}
For all $u\in\mathbb{R}^{I}_+$, let $W(u)\in[0,1]^I$ be a vector-valued function where the $i$-th element,
$W_i(u)$ is equal to $1_{i\in S}\alpha_iw_i(u_i)+1_{i\in S_0}h(i)$. Let $A$ be a $I\times I$ substochastic matrix where its $ij$-entry is equal to $a_{ij}=(1-1_{i\in S_0})(1-1_{i\in S}\alpha_i)p_{ij}$.
The solution $x^*=[x_1^*,\ldots,x_I^*]$ of the fixed-point equation, is given by:
\begin{equation}
x^*=(Id-A)^{-1}W(u),
\end{equation}
with $Id$ being the identity matrix with appropriate dimension depending on the context.

\textbf{Optimization problem: }The goal of the planner is to maximize the sum of the valuations when $k$ goes to infinity, i.e., $\sum_{i\in \SA}x_i^*$, by controlling $u_i$, under the resource constraint $\sum_{i\in S} u_i \leq M$.
Here $0 < M < |S|$ is a prescribed bound.
Equivalently, the objective of the planner is to find $u^*=(u_i^*)_{i\in S}$, the solution of the following optimization problem:
\begin{equation}
u^*=\text{arg max}_{u_i\in\mathbb{R}_+,\forall i\in S} 1^T(Id-A)^{-1}W(u),
\end{equation}
subject to:
\begin{equation}
\sum_{i\in S} u_i\leq M. \label{perstageconstraint}
\end{equation}
To compute $u^*$, we can use the gradient descent algorithm:
\begin{equation}\label{eq:classical gradient algorithm}
u_i(k+1) = \Gamma\Big(u_i(k) +  \frac{1}{k}1^T(I-A)^{-1}\frac{\partial}{\partial u_i}W(u(k))\Big),
\end{equation}
with $\frac{\partial}{\partial u_i}W(u(k))=[1_{j\in S}\alpha_j\frac{\partial w_j}{\partial u_i}(u_j)]_{j\in\SA}$. Our object is to do so in a data-driven manner using ideas from reinforcement learning and MCMC. In this paper we assume that the matrix $P$ is \textit{unknown}. The \textit{known parameters} are the set of agents $\mathcal{S}$, the vectors $\alpha :=[\alpha_i]_{i\in S}, \ h :=[h_i]_{i\in S_0}$, the functional vector $w=[w_i]_{i\in S}$ and the budget $M$. The  matrix $P$ is \textit{unknown}. At each time step $k\in\mathbb{N}_+$, the planner chooses a vector $u(k)=[u_i(k)]_{i\in S}$, then simultaneously, agent $i$ is activated w.p. $q_i$, and  polls agent $j$ probabilistically as described earlier and observes her opinion. The planner \textit{observes} this communication between $i$ and $j$. The objective of the planner is to find an algorithm such that $\lim_{k\rightarrow+\infty} u(k)=u^*$.

\section{Reinforcement learning scheme}\label{sec: first algo}
\subsection{Equivalent controlled Markov chain}
Consider an $\SA$-valued controlled Markov chain $\{Y_n\}$ with controlled transition probabilities
 \begin{eqnarray*}
q_{ij}(u)
&:=& p_{ij}, \ \ \ \ \ \ \ \ \ \ \ \ \ \ \ \ \ \ \ \ \ \ \ \ \ \ \ \ \ i \in S_1 \cup S,\\
&:=& \delta_{ij}, \ \ \ \ \ \ \ \ \ \ \ \ \ \ \ \ \ \ \ \ \ \ \ \ \ \ \ \ \ i \in S_0.
\end{eqnarray*}
Here $\delta_{ij}$ is the Kronecker delta. Thus the states in $S_0$ are absorbing states. Also note that the transition probabilities are independent of the control choice, which affects only the running reward. Let
$\tau := \min\{n \geq 0 : Y_n \in S_0\}$ denote the first passage time to $S_0$. We associate with a state-control pair $(i,u_i)$  an instantaneous cost $w_i(u)$ and a state-dependent discount factor $(1 - \alpha_i)$. Note that $\alpha_i=0$ and $w_i(u)=0$ for all $i\in S_0\cup S_1$.
Suppose $u_i(k) = v(i)$ for some $v: S \mapsto [0,1]$, i.e., a `stationary Markov policy' in Markov decision theoretic parlance \cite{BorkarCMDP}. Let $\bar{x}_i(k) := E[x_i(k)] \ \forall \ i,k$. Then $\{\bar{x}_i(k)\}$ satisfy the dynamics
\begin{eqnarray*}
\bar{x}_i(k+1) &=& \alpha_iw_i(u_i(k)) +   (1 - \alpha_i)\sum_{\ell}p_{i\ell}\bar{x}_\ell(k)), \; i \in S, \\
\bar{x}_i(k+1) &=& \sum_{\ell}p_{i\ell}\bar{x}_\ell(k)), \ \ \ \ \ \ \ \ \ \ \ \ \ i \in S_1.
\end{eqnarray*}
The matrix $P := [[p_{ij}]]_{i,j \in S\cup S_0}$ is substochastic, hence the above linear system of equations is stable. Then as $k \uparrow \infty$,
$$\bar{x}_i(k)\to \bar{x}_i(\infty),$$
where $\bar{x}(\infty)$ satisfies the equation
\begin{eqnarray*}
\bar{x}_i(\infty) &=& \alpha_iw_i(u_i(k)) +   (1 - \alpha_i)\sum_{\ell\in\SA}p_{i\ell}\bar{x}_\ell(\infty),\; i \in S, \\
\bar{x}_i(\infty) &=& \sum_{\ell\in\SA}p_{i\ell}\bar{x}_\ell(\infty), \ i \in S_1.
\end{eqnarray*}
This equation can be rewritten as:
\begin{eqnarray*}
\bar{x}_i(\infty) &=& \alpha_iw_i(u_i(k)) +   (1 - \alpha_i)\sum_{\ell\in S\cup S_1}p_{i\ell}\bar{x}_\ell(\infty)\\
&&+\sum_{\ell'\in S_0}p_{i\ell'}h(\ell'), \ i \in S, \\
\bar{x}_i(\infty) &=& \sum_{\ell\in S\cup S_1}p_{i\ell}\bar{x}_\ell(\infty)+\sum_{\ell'\in S_0}p_{i\ell'}h(\ell'), \ i \in S_1.
\end{eqnarray*}
By standard `one step analysis', one sees that $\bar{x}_i(\infty)$ has the representation
\begin{eqnarray*}
\bar{x}_i(\infty) &:=& E_i\Big[\sum_{m=0}^{\tau - 1}\left(\prod_{k=0}^{m-1}(1 - \alpha_{Y_k})\right)\alpha_{Y_m}w_{Y_m}(u(m)) \\
&&+ \ \left(\prod_{k=0}^{\tau}(1 - \alpha_{Y_k})\right)h(Y_{\tau})\Big].
\end{eqnarray*}
This suggests that we can view the opinion dynamics as the value iteration for evaluating a fixed stationary Markov policy $v(\cdot)$ for  the controlled Markov chain $\{Y_n\}$, the objective being
$\sum_{i \in S\cup S_1}\bar{x}_i(\infty)$. This will be recognized as a discounted reward for the \textit{stochastic shortest path problem} (`Longest path problem', to be precise, since we are maximizing a reward rather than minimizing a cost. The equivalent stochastic shortest path problem corresponds to viewing $-w_i(\cdot)$ as a running cost function and $-h(\cdot)$ as the terminal cost.)

We do, however, have an  additional  constraint (\ref{perstageconstraint}).
 This is hard to incorporate in a Markov decision process as a constraint on controls, because it couples actions across different states in a manner unrelated to the dynamics (i.e., without regard to, e.g., how often they are visited). This puts it beyond the reach of traditional dynamic programming based computations such as value or policy iteration, or linear programming version of the dynamic program. Therefore we treat this as a parametric optimization problem over the parameters $u_i$'s instead of as a control problem - this will become apparent from the algorithms we propose.  The `uncontrolled' but parameter-dependent  `dynamic programming'\footnote{`one step analysis', to be precise} equation  is given by standard arguments, as the linear system
\begin{eqnarray}
V(i) &=& \alpha_iw_i(u_i) +  (1 - \alpha_i)\sum_jp_{ij}V(j), \ \ \   i \in S, \label{DP1} \\
V(i)&=& \sum_jp_{ij}V(j), \   i \in S_1,\ V(i)= h(j), \  i \in S_0. \label{DP3}
\end{eqnarray}
\subsection{First algorithm}
Let $k\in\mathbb{N}_+$ be $k^{th}$ time an agent polls another.
A gradient based  learning scheme for this problem is as follows.
Let
$$
I\{Y_n = i\} =\left\{
\begin{array}{l}
= 1 \ \ \mbox{if} \ Y_n = i, \\
= 0, \ \ \mbox{if} \   Y_n \neq i, \\
\end{array}\right.\nu(i,n) := \sum_{m=0}^nI\{Y_m = i\}.
$$
for $n \geq 0$. Then $\nu(i,n), n \geq 0$, can be interpreted as a `local clock' at agent $i$, counting its own number of updates till `time' (i.e., the overall iterate count) $n$.
Pick stepsize sequences $\{a(k)\}, \{b(k)\} \subset (0, \infty)$ such that
\begin{eqnarray}
\sum_ka(k) &=& \sum_kb(k) = \infty, \; \sum_k(a(k)^2 + b(k)^2) < \infty, \nonumber \\
 \frac{b(k)}{a(k)} &\to& 0. \label{2tscale}
 \end{eqnarray}
  We shall also make the following additional assumptions on $\{a(k)\}$:
  \begin{enumerate}
  \item $a(n + 1) \leq a(n)$ from some $n$ onwards;

\item there exists $r \in (0, 1)$ such that
$\sum_na(n)^{1+q} < \infty, \  q \geq r$;

\item for $x \in (0, 1)$,
$\sup_n\left(\frac{a([xn])}{a(n)}\right) < \infty$,
where $[\cdots]$ stands for the integer part of `$\cdots$';

\item for $x \in (0, 1)$ and $A(n) := \sum_{m=0}^na(i)$,
$\lim_{n\uparrow\infty}\left(\frac{A([yn])}{A(n)}\right) = 1$
uniformly in $y \in [x, 1]$.
  \end{enumerate}
These conditions are satisfied, e.g., by the popular stepsize $a(n) = \frac{1}{n+1}, n \geq 0$.

The algorithm then is as follows. For $k \geq 0, i, j \in S$, do: for a prescribed state $i_0$,
\begin{eqnarray}
\Psi_{ij}(k+1) &=& \Psi_{ij}(k) + a(\nu(i,k))I\{Y_k = i\}\times \nonumber \\
&& \Big[\alpha_iw_i'(u_i(k))\delta_{ij} + ( 1 - \alpha_i)\Psi_{Y_{k+1}j}(k) \nonumber \\
&& - \ \Psi_{i}(k)\Big], \label{grad} \\
u_i(k+1) &=& \Gamma(u_i(k) +  b(k)\sum_j\Psi_{ji}(k)), \label{ascent} \\
\Psi_{ij}(k) &=& 0, \ \ \ \ \ \ \ \ \ \ \ \ \ \ \ \ \ \ \ \ \ i \in S_0. \label{psibdry}
\end{eqnarray}
Here $\Gamma(\cdot)$ is the projection onto the simplex
$$\{x = [x_1, \cdots, x_{|S|}]^T : x_i \geq 0 \ \forall i, \ \sum_ix_i \leq M\}.$$
This is a gradient-based reinforcement learning scheme which is better suited for our purposes than, e.g., the classical Q-learning scheme of \cite{ABB}. This is because it allows us to treat the optimization over control parameters as parametric optimization which can handle the constraint (\ref{perstageconstraint}). The explanation of this scheme is as follows:
\begin{itemize}
\item The iteration (\ref{grad}) estimates the partial derivatives $\frac{\partial V(i)}{\partial u_j}$ by $\Psi_{ij}(k), k \geq 0$. This is arrived at by considering the constant policy dynamic programming equation
\begin{equation}
\widetilde{V}(i) = \alpha_iw_i(u_i) + (1 - \alpha_i)\sum_jp_{i\ell}\widetilde{V}(\ell), \label{coPo}
\end{equation}
for $i \in S$.
Differentiating both sides w.r.t.\ $u_j$, we see that $\Phi_{ij} := \frac{\partial V(i)}{\partial u_j}$ satisfy
\begin{equation}
\Phi_{ij} = \alpha_iw_i'(u_i)\delta_{ij}  + (1 - \alpha_i)\sum_{\ell}p_{i\ell}\Phi_{\ell j}, \label{grad0}
\end{equation}
 for $i \in S$, with $\Phi_{ij} = 0$ for $i \in S_0$.
  The iteration (\ref{grad}) then is the stochastic approximation scheme to solve this equation.

\item Iteration (\ref{ascent}) operating on a slower time scale (in view of (\ref{2tscale})), constitutes a stochastic gradient ascent. That is, (\ref{ascent}) is a stochastic gradient ascent over the control variables which takes the outputs $\{\Psi_{ij}(k)\}$ of (\ref{grad}) as estimates of the relevant partial derivatives and summing them over the first index, generates an estimate of the corresponding partial derivative of the reward itself. In turn, the application of the projection $\Gamma( \cdot )$  makes it a projected stochastic gradient scheme which also imposes the constraint (\ref{perstageconstraint}).

\end{itemize}

The chain $\{Y_n\}$, however, is an imaginary object. To map this scheme back to our original framework, let $Z_n$ be the index of the agent that updated its valuation at time $n$. In other words, it is the $Z_n$-th component $x_{Z_n}(n)$ that got updated at time $n$, the rest were left unperturbed. Also suppose that this was done by the $Z_n$-th agent by polling a neighbor $\widetilde{Z}_n$  according to the transition probabilities $q_{z_n\cdot}$ defined above.

The  iteration (\ref{grad}) of the above scheme can then be written for our original framework as
\begin{eqnarray}
\lefteqn{\Psi_{ij}(k+1) = \Psi_{ij}(k) + a(\nu(i,k))I\{Z_k = i\}\times } \nonumber \\
&&\Big[(\alpha_iw_i'(u_i(k))\delta_{ij}
 + ( 1 - \alpha_i)\Psi_{\widetilde{Z}_kj}(k) - \Psi_{ij}(k)\Big]. \label{grad1}
\end{eqnarray}
The third  iteration, i.e., (\ref{ascent}), remains unaltered, as do the boundary conditions for the $\Psi_{i}$'s.

Now that we no longer have to think of $Z_k, \tilde{Z}_k$ as realizations of a single trajectory of a Markov chain, we can generalize this further and let $Z_k$ be a subset of $S$. For each $i \in Z_k$, we generate a random variable $\tilde{Z}^i_k$ according to the probability distribution $q_{i\cdot}$ above. The iteration (\ref{grad2}) then gets replaced by
\begin{eqnarray}
\lefteqn{\Psi_{ij}(k+1) = \Psi_{ij}(k) + a(\nu(i,k))I\{i \in Z_k\}\times } \nonumber \\
&&\Big[(\alpha_iw_i'(u_i(k))\delta_{ij}
 + ( 1 - \alpha_i)\Psi_{\widetilde{Z}^i_kj}(k) - \Psi_{i}(k)\Big]. \label{grad2}
\end{eqnarray}
In particular when $Z_n = S$, it is a completely synchronous iteration. Note that $\nu(i,n)$ has to be defined now as $\nu(i,n) := \sum_{m=0}^nI\{i \in Z_m\}.$

\subsection{Stochastic gradient scheme}

Along the line of the previous algorithm, we define a new algorithm, where instead of having a biased but consistent estimator of the gradient, we can derive a sampling scheme that will provide, at each iteration, a unbiased estimator of the gradient. Recall the $Z_n, \tilde{Z}_n$  defined in the preceding section. There we had considerable freedom in choosing how $Z_n$ is generated, the key requirement was that $\tilde{Z}_n$ should have the prescribed conditional law given $Z_n$. This is because the algorithm at each step calls for a single transition executed according to the given transition matrix. That is, one has to generate a pair of random variables with the conditional law of the latter given the former completely specified and the (marginal) law of the former  having full support at each step. In the algorithm we propose below (with its natural extension), however, we require at each step a path of random duration in $\SA$. Generating pairs $(Z_n, \tilde{Z}_n)$ as before does not provide that. Hence unlike the previous scheme with full observations, we now need a probing mechanism. Thus we define $Z_n$ as before and do (for each fixed $n$): Let $\delta_{\cdot \cdot}$ denote the Kronecker delta.
\begin{enumerate}
\item  For each $i\in S$, set $m = 0$ and set $Y_{j0} = Z_n = j$ (say). Here $j \in \SA\backslash S_0$ can be picked uniformly at random.  Initialize $\xi_{ji}(n) = 0$.
\item With probability $\alpha_{Y_{j0}} = \alpha_{j}$ ($\alpha_{j}=0$ if $j\in S_1$),  stop and set  $\xi_{ji}(n) \rightarrow \xi_{ji}(n) + \delta_{ji}$.
If not,
\item with probability $(1 - \alpha_{Y_{j0}})p_{Y_{j0}k}$, continue by setting  $Y_{j1}= k, \ \xi_{ji}(n) \rightarrow \xi_{ji}(n)$.
\item At step $m$, stop if $Y_{jm} \in S_0$. If not, stop with probability $\alpha_{Y_{jm}}$ and set $\xi_{ji}(n) \rightarrow \xi_{ji}(n) + \delta_{Y_{jm}i}$, or else continue  with probability $(1 - \alpha_{Y_{jm}})$ by setting $Y_{j(m+1)} = \ell$ with probability $(1 - \alpha_{Y_{jm}})p_{Y_{jm}\ell}$.
\item Repeat 4) above for $m \geq 1$ till stopping.
\item Perform the following gradient descent step:
\begin{equation}\label{sgd}
u_i(n+1) = \Gamma\left(u_i(n) +  a(n)w_{i}'(u_{i}(n))\sum_i\xi_{ij}(n)\right). \\
\end{equation}
\end{enumerate}
An alternative scheme is:
\begin{enumerate}
\item  For each $i\in \SA$, and for each $j\in S$, set $m = 0$ and $Y_{i0}  = i$, kept fixed for this run. Initialize $\zeta_i = 1$.  Set
$$\xi_{ij} \rightarrow \xi_{ij} + \zeta_i\delta_{Y_{i0}j}\alpha_{Y_{i0}}.$$
Continue by setting  $Y_{i1}= k$ with probability $p_{Y_{i0}k}$.
\item At step $m$, stop if $Y_{im}  \in S_0$. If not,  set
\begin{equation*}
\zeta_i \rightarrow \zeta_i(1 - \alpha_{Y_{i(m-1)}}), \;\xi_{ij} \rightarrow \xi_{ij} + \zeta_i\delta_{Y_{im}j}\alpha_{Y_{im}},
\end{equation*}
and continue  by setting $Y_{i(m+1)} = \ell$ with probability $p_{Y_{im}\ell}$.

\item Repeat 2) above for $m \geq 1$ till stopping. Freeze $\xi_{ij}$ on stopping and label it $\xi_{ij}(n)$.
\item Perform the following gradient descent step:
\begin{equation}\label{sgd2}
u_i(n+1) = \Gamma\left(u_i(n) +  a(n)w_{i}'(u_{i}(n))\sum_j\xi_{ji}(n)\right). \\
\end{equation}
\end{enumerate}

By construction, for the two sampling schemes, $w_{i}'(u_{i}(n))E\left[\sum_\ell\xi_{i\ell}(n)\right]$, is the solution of the linear system \eqref{grad0}. Therefore, the previous scheme will converge to the optimal $u^*$ as long as the variance of  $\xi_{ij}(n)$ is bounded for all $k$ \cite{BorkarBook}. For the stochastic gradient iterate (\ref{sgd2}), a good step-size in this context is $ a(n)=A/(\lceil\frac{n}{M}\rceil)$ for some $A > 0$ and  $M \geq 1$.

\subsection{An alternative learning scheme}

The problem with the above scheme is that it involves all agents in $\SA$, which may lead to a curse of dimensionality. Worse, it requires that all communications between agents be observed. It makes sense to assume that only a few agents can be monitored. These should include in particular those in $S$. Without any loss of generality, we assume that only the updates of agents in $S$ are observed. The algorithm we propose next and its analysis extend easily to the case when a few uncontrolled agents are also observed (by using, e.g., the trivial device of setting $\alpha_i \equiv 0$ for such agents). Then it also makes sense that we should treat $S^* := S\cup S_0$ as our effective state space for the algorithm. By analogy to the above stochastic shortest path formulation, consider an $\SA$-valued Markov chain $\{Y_n\}$ with transition probabilities $\{p_{ij}\}$. If we restrict $\{Y_n\}$ to $S^*$, it means that we observe only $\{Y_{T_n}\}$ where $T_n, n \geq 0,$ are the successive return times of $\{Y_n\}$ to $S^*$ define recursively by
\begin{eqnarray*}
T_0 &:=& \min\{m \geq 0 : Y_m \in S^*\}, \\
T_{n+1} &:=& \min\{m > T_n : Y_m \in S^*\}, \ n \geq 0.
\end{eqnarray*}
The chain eventually gets absorbed into $S_0$ as before. Strictly speaking, if we keep track of the $T_n$'s as well, it is a \textit{semi-Markov} process. Exercising control only when the chain is in $S^*$ leads to a supervisory control problem as in \cite{Varaiya}, albeit with a different reward structure compared to theirs. Nevertheless, we do not need to view it in this manner. This is because our controlled Markov chain is an artifact, the actual process is the simple averaging or `gossip' dynamics. Thus the actual values of $T_n$'s are irrelevant for us and we can work with the chain $Y^*_n := Y_{T_n}, n \geq 0$.  Let $q^*(j | i), i, j \in S^*,$ denote the probability that $Y^*_{n+1} = j$ given $Y^*_n = i$. It is of the form
$$q^*(j|i) = p_{ij} + \sum_{j \neq \ell \in S'}p_{i\ell}\varphi(j|\ell)$$
where $\varphi(j | \ell) := P(Y^*_{\zeta} = j | Y^*_0 = \ell)$, for $\zeta := \min\{n \geq 0 : Y^*_n \in S\}$. In particular, $\varphi( \cdot | \cdot )$  is independent of the control choice $u$.  After the chain leaves state $i$, it does not hit any other controlled state before hitting another state ($j$ above) in $S^*$, so the associated running cost is $\alpha_iw_i(u)$ as before.  We now consider the restricted reward $\sum_{i\in S}x_i(\infty)$ which is \textit{not} the same as the original, so this is an approximation. The advantage of this reward is that it is expected to be positively correlated with the full reward, i.e., increase in the former should lead to increase in the latter. More importantly,  it depends only on observed quantities. This passage is purely heuristic and avoids in particular having to contend with the full complications of the `partial observations' framework.  The associated (constant policy) dynamic programming equation is then given by
\begin{eqnarray}
V(i) &=& \alpha_iw_i(u_i) + (1 - \alpha_i)\big(p_{ij} +  \nonumber\\
\ && \ \sum_{j, \ell \in S' : j \neq \ell}p_{i\ell}\varphi(j|\ell)\big)V(j),
\label{newDP1} \\
V(i) &=& h(i), \ \ \ \ i \in S_0. \label{newDP2}
\end{eqnarray}

One could write down a reinforcement learning scheme for approximate solution of (\ref{newDP1})-(\ref{newDP2}) along the lines of the preceding subsections, but the situation is much more difficult here. The problem is similar to the one faced in the stochastic gradient scheme above.
We require a path from one state in $S^*$ to another, passing through a possibly nonempty set of unobserved states in $\SA\backslash S^*$. Again, generating pairs $(Z_n, \tilde{Z}_n)$ as before does not provide that.  We now need a probing mechanism. Thus we define $Z_n$ as before, but when node $Z_n = i \in S^*$ polls a neighbor $i_1\in\SA$, it passes to $i_1$ a time-stamped token tagged with $i$. The node $i_1$, if not in $S^*$, does likewise, but retaining the original tag and time stamp. This continues till the token reaches some $j \in S^*$. Then set $\tilde{Z}_n = j$.
The corresponding reinforcement learning scheme now becomes
\begin{eqnarray}
\Psi_{i}(k+1)
 &=& \Psi_{i}(k) + a(\nu(i,k))I\{i \in Z_k\}\times \nonumber \\
&& \Big[(\alpha_iw_i'(u_i(k)) + ( 1 - \alpha_i)\Psi_{\widetilde{Z}^i_k}(k) \nonumber \\
&& - \ \Psi_{ij}(k)\Big],  \label{grad21} \\
u_i(k+1) &=& \Gamma\Big(u_i(k) +  b(k)\Psi_{i}(k)\Big), \label{ascent2} \\
\Psi_{i}(k) &=& 0, \ \ \ \ \ \ \ \ \ \ \ \ \ \ \ \ \ \ \ \ i \in S_0. \label{psibdry3} 
\end{eqnarray}

\section{Convergence analysis}\label{sec: Convergence analysis}

The convergence analysis of the first scheme goes along standard lines, essentially piecing together known facts from the theory of two time scale and distributed asynchronous stochastic approximation. With this in mind, we only sketch it in outline. To begin with, note that condition (\ref{2tscale}) implies that the iterates (\ref{ascent2}) move on a slower, in fact asymptotically negligible,  time scale compared to (\ref{grad2}). Hence they can be viewed as quasi-static, i.e., $u_i(k) \approx u_i \ \forall i$, for purposes of analyzing (\ref{grad2}) (\cite{BorkarBook}, Section 6.1).
Then (\ref{grad2}) constitutes a stochastic approximation scheme to estimate the partial derivatives of $V^*$ w.r.t.\ the $u_i$'s by solving (\ref{grad0}), which has a unique solution. Its convergence to this solution follows from the theory of asynchronous stochastic approximation developed in \cite{asyn}, wherein the conditions we have imposed on $\{a(n)\}$  play a crucial role.

But this is under the assumption that $u_i(k) \approx u_i \ \forall i$, whereas in reality the $u_i(k)$'s are changing on a slower time scale. Thus what the foregoing entails in reality is that
$$\Psi_{ij}(k) - \frac{\partial V^*(i)}{\partial u_j}\Big\vert_{u_\cdot = u_\cdot(k)} \to 0$$
a.s.\ $\forall \ i,j$, i.e., $\Psi_{ij}$'s track the corresponding partial derivatives of $V^*$ with an asymptotically negligible error, as desired. Thus (\ref{ascent2}) is a legitimate stochastic gradient ascent scheme. We need the following lemma.

%
\noindent \textbf{Lemma 1} The solution $V(\cdot)$ of the constant policy dynamic programming equation (\ref{newDP1})-(\ref{newDP2})
is componentwise concave and continuous in the variables $\{u_i\}$.

\noindent \textbf{Proof (Sketch)} This follows by considering the associated constant policy value iteration and using induction, along with the fact that pointwise limits of concave functions are concave and uniform limits of continuous functions are continuous. The details are routine, see, e.g., \cite{Agar}. \hfill $\Box$


Our main result then is the following:

\noindent \textbf{Theorem 1} The above learning policy is asymptotically optimal, a.s.\\

\noindent \textbf{Proof} This is immediate from the fact that the projected stochastic gradient ascent for a concave function on a compact interval converges to the set of its global maxima a.s.\ (see, e.g., \cite{BorkarBook}, Chapter 10). \hfill $\Box$

The stochastic gradient scheme above, by virtue of (\ref{grad0}), is already of the form
\begin{eqnarray*}
\lefteqn{u_i(n+1) = u_i(n) + } \\
&&a(n)I\{Z_n = i\}\left[\frac{\partial V(i)}{\partial u_i}(u(n)) + M_i(n+1)\right],
\end{eqnarray*}
where $M(n) := \left[M_1(n), M_2(n), \cdots \right]^T$ is a martingale difference sequence. That is, it is a classical asynchronous stochastic gradient scheme with a.s.\ convergence to a local miaximum, which is also a global miaximum by concavity of $V(i)$,  under reasonable conditions on $\{M(n)\}$ -- see Chapter 10 of \cite{asyn}.

Finally, the `alternative scheme' above based on observing few nodes is of the same form as the reinforcement learning scheme above and is analyzed exactly the same way.
\section{ A more general model}\label{sec: a more general model}

We can also consider the situation where $\alpha_i$'s depend on the control choice $u_i$ at $i \in S$.  We shall illustrate the changes for the second, i.e., the improved learning scheme above, the situation for the first scheme being completely analogous. Thus the `dynamic programming equations' become
\begin{eqnarray*}\nonumber
V(i) &=& \alpha_i(u_i)w_i(u_i) + \\
&&(1 - \alpha_i(u_i))\big(p_{ij} +  \sum_{\ell \in S'}p_{i\ell}\varphi(j|\ell)\big)V(j),  \\
V(i) &=& h(i), \ \ \ \ i \in S_0,
\end{eqnarray*}
and the corresponding reinforcement learning scheme is
\begin{eqnarray}\nonumber
V_i(k+1) &=& V_i(k) + a(\nu(i,k))I\{i \in Z_k\}\times \\\nonumber
&&\Big[\alpha_i(u_i(k))w_i(u_i(k)) + \\\label{eq: approx grad gm 1}
&&(1 - \alpha_i(u_i(k)))V_{\widetilde{Z}_k^i}(k) - V_i(k)\Big],  \\\nonumber
\Psi_{ij}(k+1) &=& \Psi_{ij}(k) + a(\nu(i,k))I\{i \in Z_k\}\times \\\nonumber
&&\Big[(\alpha_i(u_i(k))w_i'(u_i(k)) + \alpha_i'(u_i(k))w_i(u_i(k)))\delta_{ij} \\
&& - \ \alpha_i'(u_i(k)))\delta_{ij}V_{\widetilde{Z}_k^i}(k)  \\\label{eq: approx grad gm 2}
&&+ \ ( 1 - \alpha(u_i(k)))\Psi_{\widetilde{Z}_k^ij}(k) - \Psi_{ij}(k)\Big],  \\\label{eq: dec gm 1}
u_i(k+1) &=& \Gamma\Big(u_i(k) +  b(k)\sum_j\Psi_{ji}(k)\Big),  \\\nonumber
V_k(i) &=& h(i),\ i \in S_0,\;\Psi_{ij}(k) = 0, \ i \in S_0.
\end{eqnarray}
The convergence analysis applies as before except for the fact that we can no longer claim concavity and convergence only to a local minimum can be guaranteed. This could be improved, e.g., by resorting to simulated annealing for the slow time scale iterates, i.e., replacing
them by
\begin{eqnarray}\nonumber
u_i(k+1) &=& \Gamma\Big(u_i(k) +  b(k)\sum_j\Psi_{ji}(k) \\\label{eq: dec gm 2}
&&+ \ \frac{C}{\sqrt{1/b(k)\log\log (c(k))}}W_{k+1}\Big),
\end{eqnarray}
where $\{W_k\}$ are IID N(0,1), $C > 0$ is a suitably chosen constant as in \cite{Gelfand}.The difference with the previous scheme is that (\ref{coPo}) gets replaced by
\begin{equation}
\widetilde{V}(i) = \alpha_i(u_i)w_i(u_i) + (1 - \alpha_i(u_i))\sum_jp_{i\ell}\widetilde{V}(\ell). \label{coPo2}
\end{equation}
Differentiating through with respect to $u_i$ in (\ref{grad0}) and after replacing $\alpha_i$ by $\alpha(u_i)$, we have the additional term $\alpha'_i(u_i)(w_i(u_i) - \sum_{\ell}p_{i\ell}\widetilde{V}(\ell))$ on the right hand side. The second and third term on the right had side of (\ref{eq: approx grad gm 2}) correspond to these additional terms. As this involves $\tilde{V}(\cdot)$ as well unlike the previous scheme which did not, one needs the additional iteration (\ref{eq: approx grad gm 1}) to estimate it, this being the stochastic approximation scheme to solve (\ref{coPo}).

For comparison purposes later on in the numerical section, we also state the iterations in the case where we know the matrix $P$ explicitly. Then the only difference would be in the update of $V_i(k)$ and $\Psi_{ij}(k)$ which will follow the following scheme:
\begin{eqnarray}\nonumber
V_i(k+1) &=& V_i(k) + a(k)\times \\\nonumber
&&\Big[\alpha_i(u_i(k))w_i(u_i(k)) + \\\label{eq: approx grad gm 3}
&&(1 - \alpha_i(u_i(k)))\sum_{l\in\SA}p_{il}V_{l}(k) - V_i(k)\Big],  \nonumber
\end{eqnarray}
\begin{eqnarray}\nonumber
\Psi_{ij}(k+1) &=& \Psi_{ij}(k) + a(k)\times \\\nonumber
&&\Big[(\alpha_i(u_i(k))w_i'(u_i(k))\delta_{ij} +  \\\nonumber
&&\Big[(\alpha_i'(u_i(k))w_i(u_i(k)) - \alpha_i'(u_i(k)))\delta_{ij}\times\\\label{eq: approx grad gm 4}
&&\sum_{l\in\SA}p_{il}V_{l}(k),  \\\nonumber
V_k(i) &=& h(i), \   i \in S_0,\;\Psi_{ij}(k) = 0, \  i \in S_0.
\end{eqnarray}

\section{Numerical experiments}\label{sec: numerical experiments}
\begin{figure*}[t]
    \centering
   \subfigure[Evolution of $u(k)$.]{\includegraphics[scale=0.3]{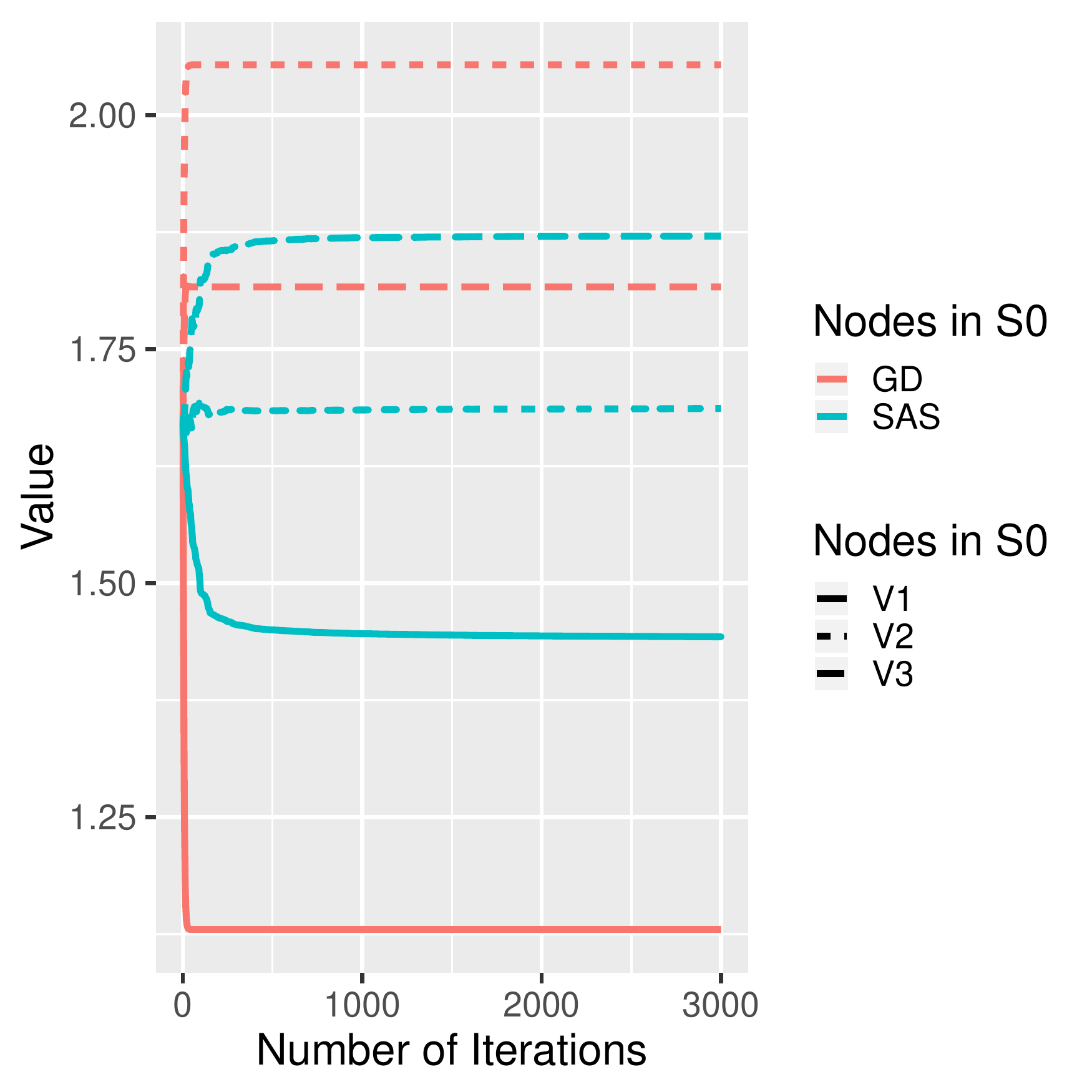}\label{fig:Plot_Convergence_1_1_imperfect}}
   \subfigure[Evolution of the relative difference between the current payoff and the optimal payoff.]{\includegraphics[scale=0.3]{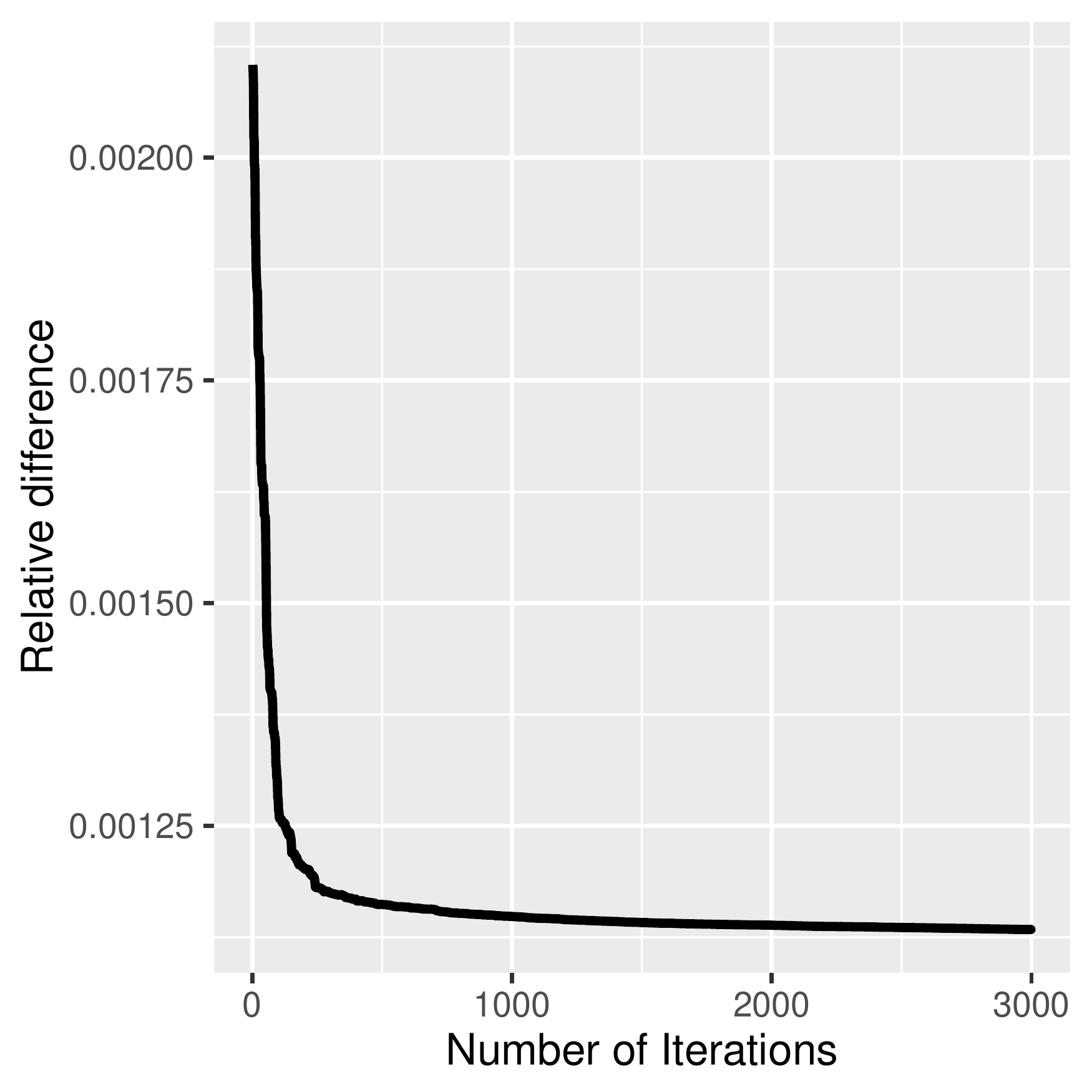}\label{fig:Plot_Convergence_1_2_imperfect}}
   \subfigure[Box-plot.]{\includegraphics[scale=0.3]{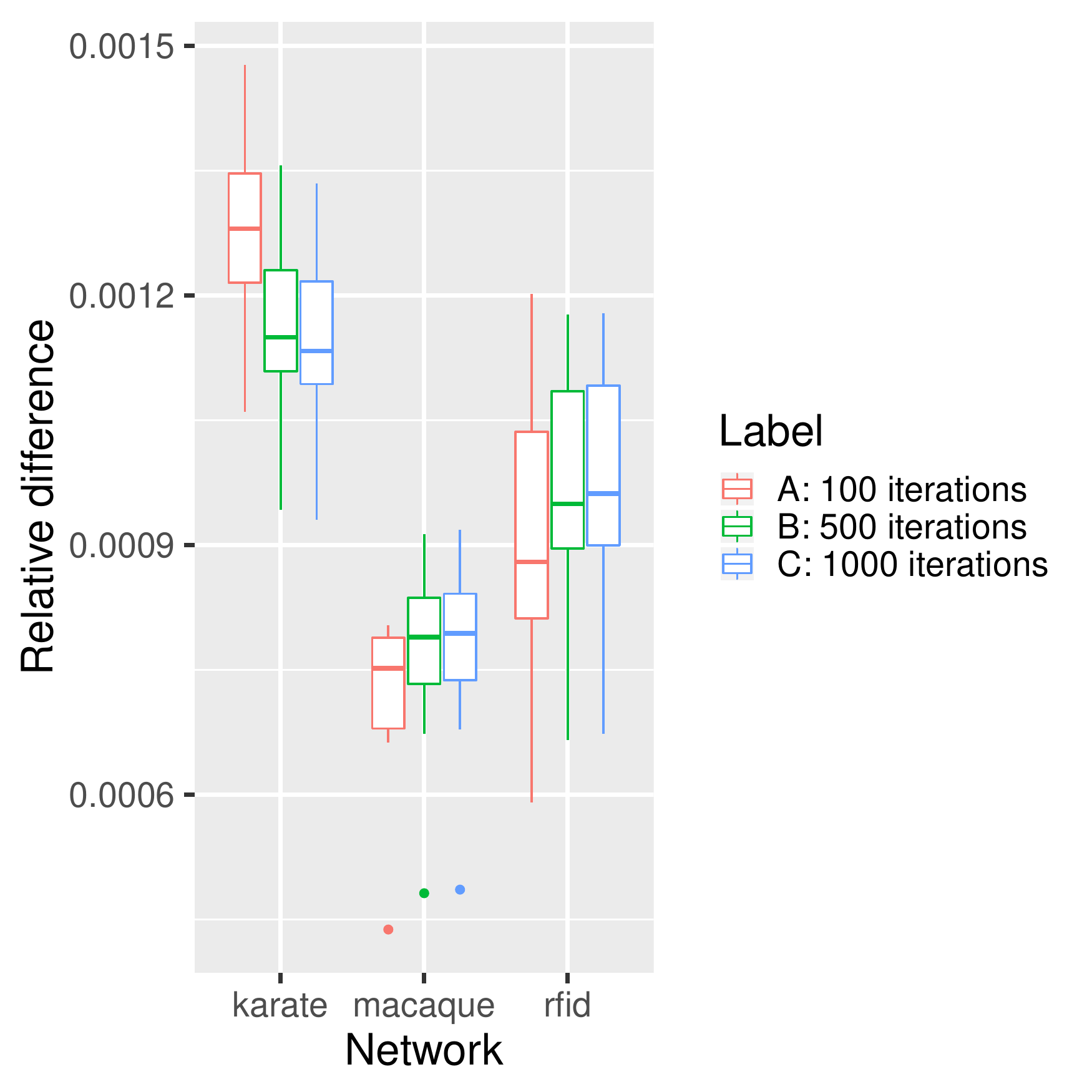}\label{fig:box_plot_imperfect}}
    \caption{Simulations for the improved learning scheme.}
\end{figure*}
We select three real-world networks for our evaluation. The three networks are Karate, Macaque, Rfid (see Table \ref{tab:networks description}) and have from 34 to 75 nodes and from 78 to 2278 links. Each network was retrieved from the R package igraphdata \cite{csardi2015igraphdata}. The numerical experiments reported here are for the synchronous case, i.e.,  all components are updated each time. The results are compared with the exact solution computed off-line using gradient descent described in \eqref{eq:classical gradient algorithm}.

\noindent \textit{Inputs:} the matrix $P$, the number of agents in each set ($S$, $S_1$, $S_0$), the upper bound in the resource constraint $M$, the number of iterations, the function $w(\cdot)$ and finally the parameters $A$, $B$ and $denom$  of our step-size functions $a(k)=\frac{A}{\lceil(1+klog(1+k))/denom\rceil}$ and $b(k)=\frac{B}{\lceil k/denom \rceil}$. 

\noindent\textit{Construction of $P$ :} Given an adjacency matrix $A$, which can be weighted or not, we transform this matrix in a stochastic matrix by dividing each row by the sum of its elements. This matrix is our communication matrix $P$.

\noindent \textit{Initial setting:} First we specify the number of agents in each set ($S$, $S_1$, and $S_0$) and then randomly allocate an agent to a given set. We assume that $\alpha_i=\alpha$ for each $i\in S_0\cup S$ and $\alpha_i=0$ for all $i\in S_1$. For each $i\in S_0$, $h(i)$ is sampled from a uniform distribution. In our simulations, $\alpha=0.6$, $M=5$, $A=0.6$, $B=0.6$, $denom=100$ and $w(x)=\frac{x}{x+0.1}$.

  \begin{table}[ht]
  \centering
  \begin{tabular}{|l||c|c|c|}
  \hline
  Network&Karate &macaque& rfid \\
  \hline
    Number of nodes & 34 & 45 & 75  \\
\hline
    Number of edges & 78 & 463 & 2278\\
\hline
  \end{tabular}
 \caption{Description of the networks}
 \label{tab:networks description}
\end{table}

\begin{figure}[h]
    \centering
  \subfigure[SAS.]
    {\includegraphics[scale=0.3]{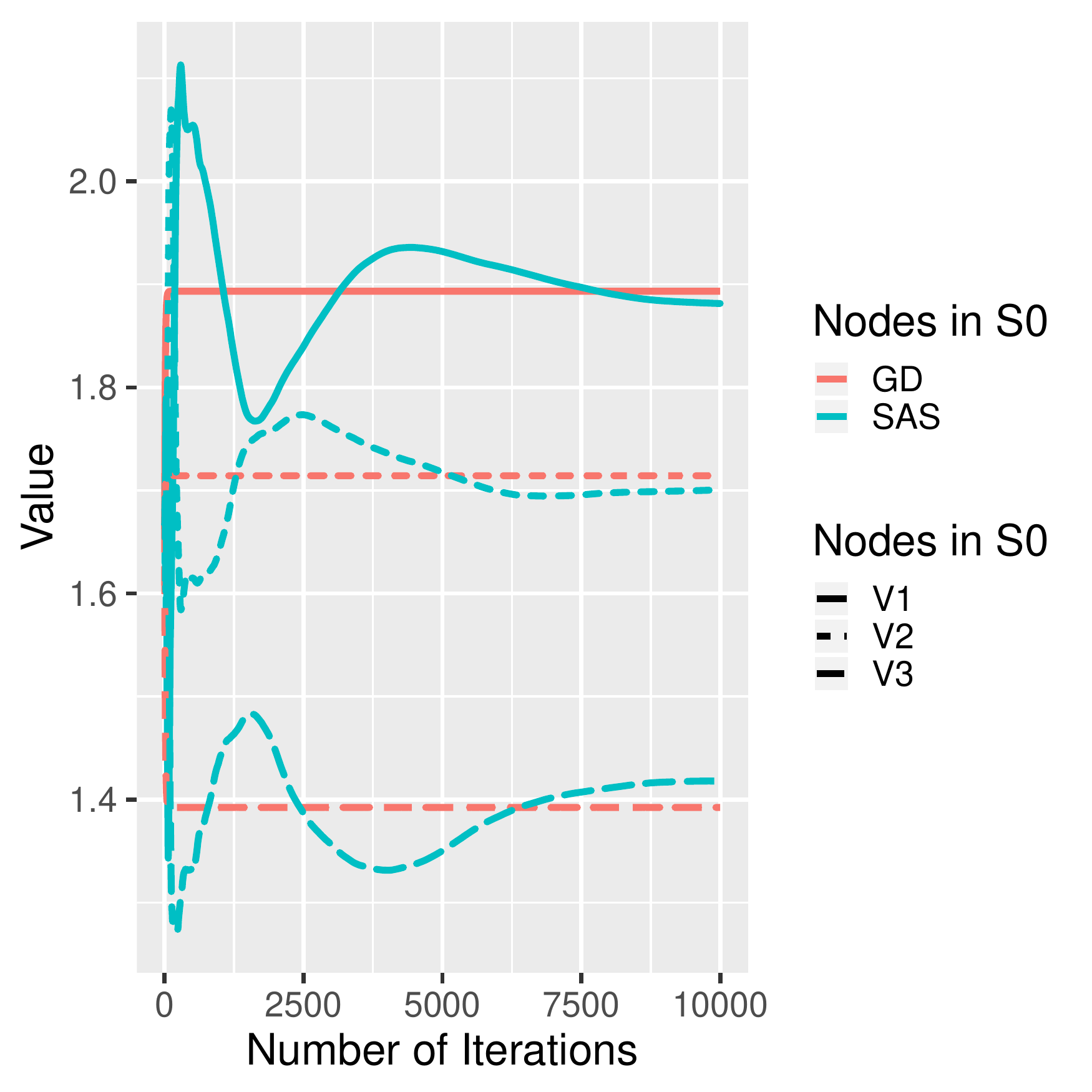}\label{fig:Plot_Convergence_1_3}}
        \subfigure[SGD.]
    {\includegraphics[scale=0.3]{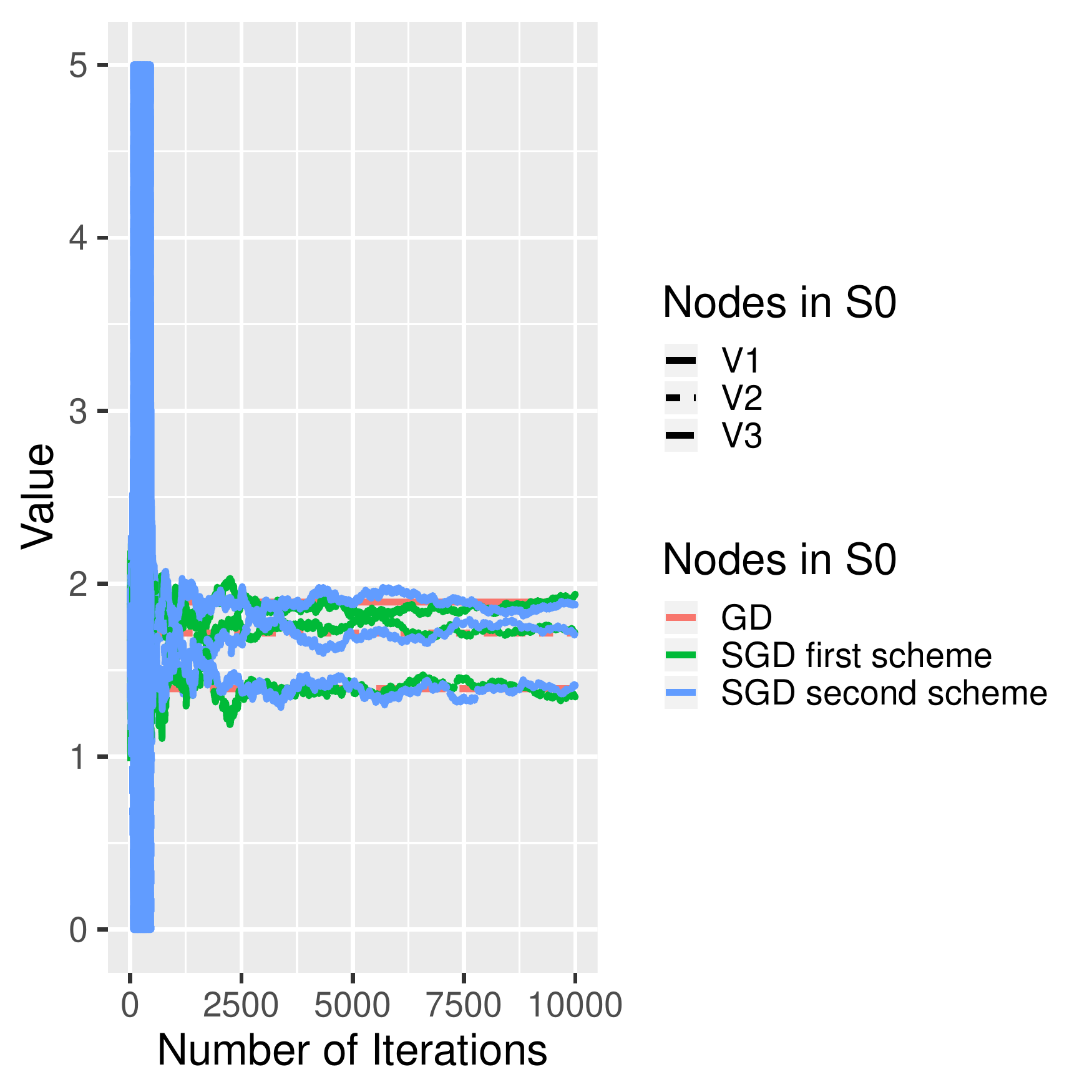}\label{fig:Plot_Convergence_1_3_40000}}
    \caption{Evolution of $u(k)$ for each algorithm.}
    {\includegraphics[scale=0.3]{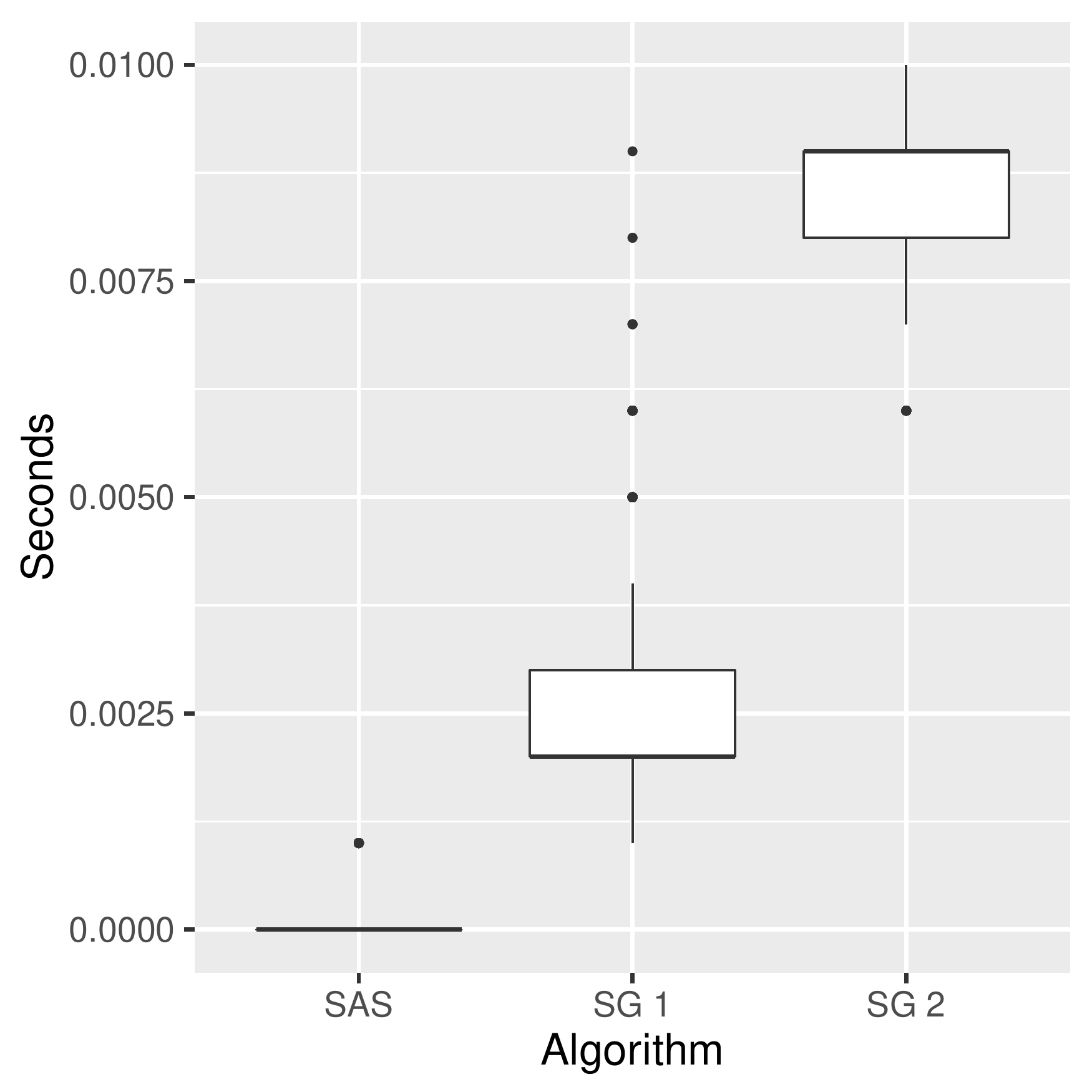}\label{fig:boxplot}}
    \caption{Boxplot for the time (in seconds) to perform one iteration of each algorithm.}
\end{figure}

\noindent\textbf{Convergence for the Karate network.} In the first numerical study, we are interested in  understanding the convergence of the stochastic approximation scheme and the stochastic gradient to the optimal strategy. We restrict this study to the Karate network. Later on, we shall extend it to the remaining networks. In Figure \ref{fig:Plot_Convergence_1_3}, the x-axis denotes the number of iterations and the y-axis captures the evolution of $u(k)$ for the stochastic approximation/reinforcement based scheme \eqref{grad}-\eqref{psibdry}. We will abbreviate the name of this scheme by SAS (for stochastic approximation scheme).  In Figure \ref{fig:Plot_Convergence_1_3_40000}, the x-axis denotes the number of iterations and the y-axis captures the evolution of $u(k)$ for the stochastic gradient (SGD) with the two sampling schemes \eqref{sgd} and \eqref{sgd2}. In the two figures, the red curve captures the evolution of $u(k)$ using the gradient descent (GD). The number of controlled agents is equal to 3. Twenty-eight agents belong to $S_1$ and three agents are in $S_0$. In Figure \ref{fig:Plot_Convergence_1_3}, before 7500 iterations, we can observe that the gradient descent algorithm already converges and the reinforcement learning scheme did not. In fact the SGD seems to converge faster (see Figure \ref{fig:Plot_Convergence_1_3_40000} after 2500 iterations). However, we observe that the variance over the iterates of SGD is higher than the SAS. The tradeoff therefore is between speed and fluctuations. Moreover we can observe in Figure \ref{fig:boxplot} that one iteration of the SAS is much faster than the ones of the two SGD algorithms. Therefore there is a clear tradeoff between the complexity of a single iteration and the number of iterations, so the latter cannot be the sole basis for comparison.

\begin{figure}[h]
  \centering
   \subfigure[SAS.]{\includegraphics[scale=0.3]{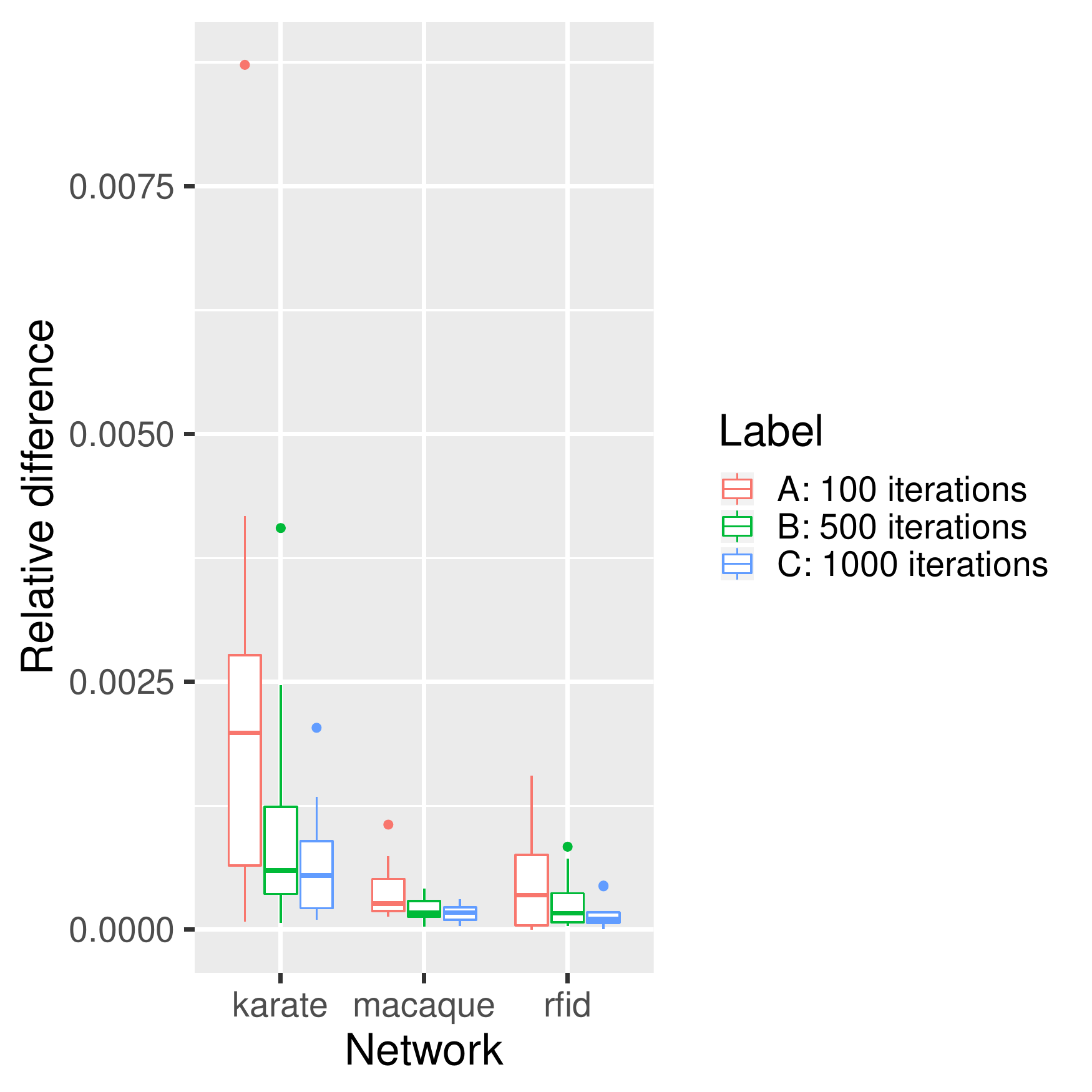}\label{fig:Box_plot_3}}
    \subfigure[SGD first scheme.]{\includegraphics[scale=0.3]{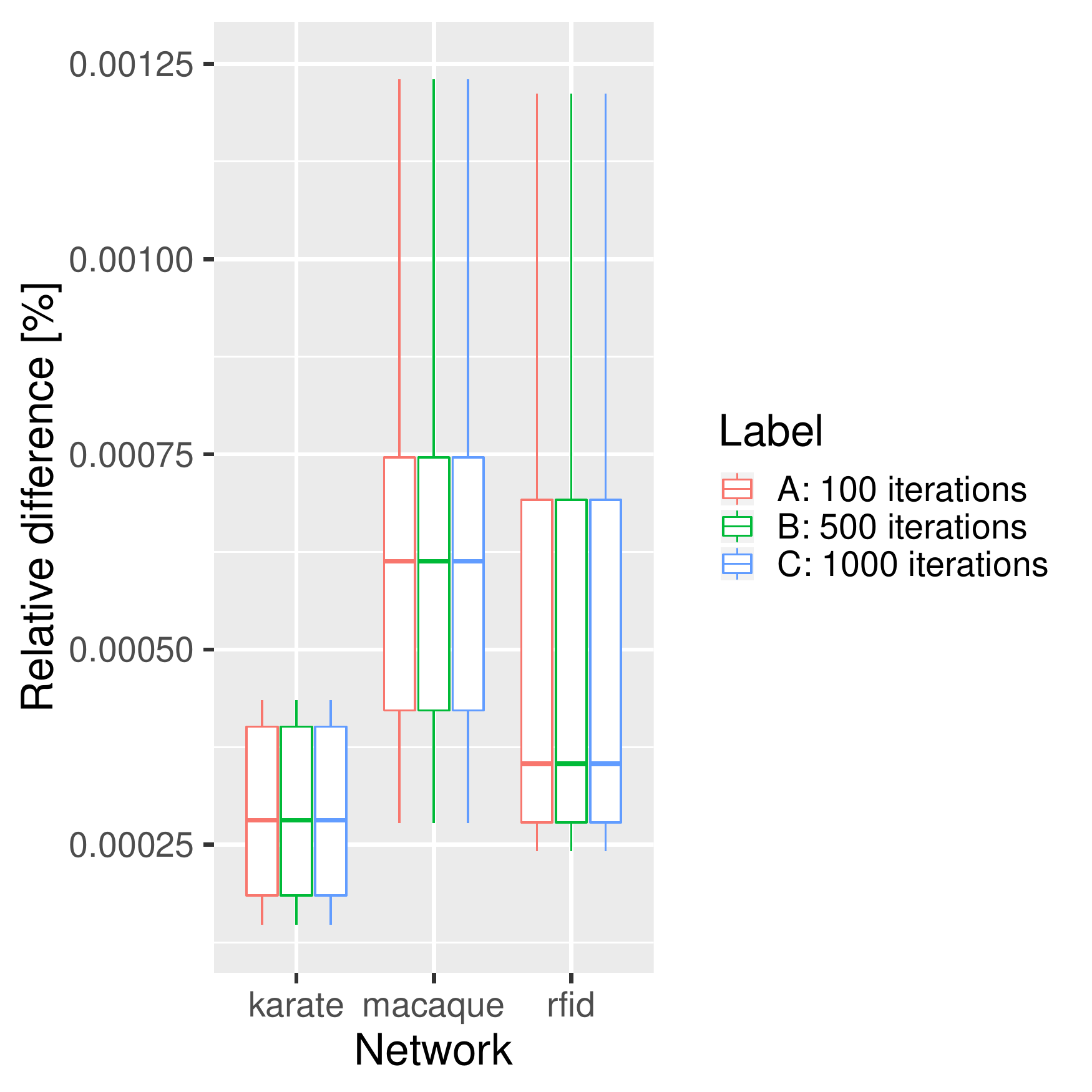}\label{fig:Box_plot_5}
}
\caption{Box-plot of the relative difference between the payoff obtained at $k$ and the optimum over 10 simulations for the stochastic approximation scheme and the stochastic gradient descent.}
\end{figure}

\noindent\textbf{Extension to other networks:} The second numerical study applies the same schemes to the other  datasets and  observes whether or not the same conclusions apply. We do not present the SGD with the second sampling scheme because the conclusion are similar. In Figure \ref{fig:Box_plot_3} and in Figure \ref{fig:Box_plot_5}, we perform 10 simulations of the stochastic approximation scheme and stochastic gradient for each network. The performance measure on $y$-axis is the relative difference between the optimal payoff and the current payoff generated by $u(k)$ at iteration $k$. For the SAS, we observe that for each network, even if we stop the stochastic approximation after 100 iterations, the third quantile will have a relative difference lower than 1\%. For each network, when we use the stochastic gradient, we note that relative difference is much lower that for the SAS. The last observation highlights the fact that when the number of iterations is low (under 1000 in this case), the SAS uses a biased estimator of the gradient compared to the stochastic gradient and therefore has lower performance.

\noindent\textbf{Study of the improved learning scheme:} In the third numerical study, we are interested in understanding how the second learning scheme compares with the first. The main difference between the two algorithms is that in the first one you have to observe the communication between all the agents and in the improved one, you can only observe a part of it. In order to be able to compare with the previous simulations, we assume the following: The set of controlled agents is the same ($S$ is the same). Only 50\% of the agents in $S_0$ and $S_1$ are observed. The results are depicted in Figures \ref{fig:Plot_Convergence_1_1_imperfect}, \ref{fig:Plot_Convergence_1_2_imperfect} and \ref{fig:box_plot_imperfect}. In Figure \ref{fig:Plot_Convergence_1_1_imperfect}, we note that the improved stochastic approximation scheme already converges after a number of iterations less than $3000$. The convergence is not to the optimal one but in this case, we can observe that in Figure \ref{fig:Plot_Convergence_1_2_imperfect}, the relative difference of the current payoff and the optimal is below 0.1\%, therefore nearly optimal. We can conclude that even if the improved stochastic approximation does not converge to the optimal $u^*$, the strategy reached is already quite good. We can observe a similar conclusion in Figure \ref{fig:box_plot_imperfect} for the remaining networks. These preliminary simulations encourage the use of the improved stochastic approximation scheme.

\begin{figure}[h]
  \centering
   \subfigure[$denom=1000$.]{
\includegraphics[scale=0.3]{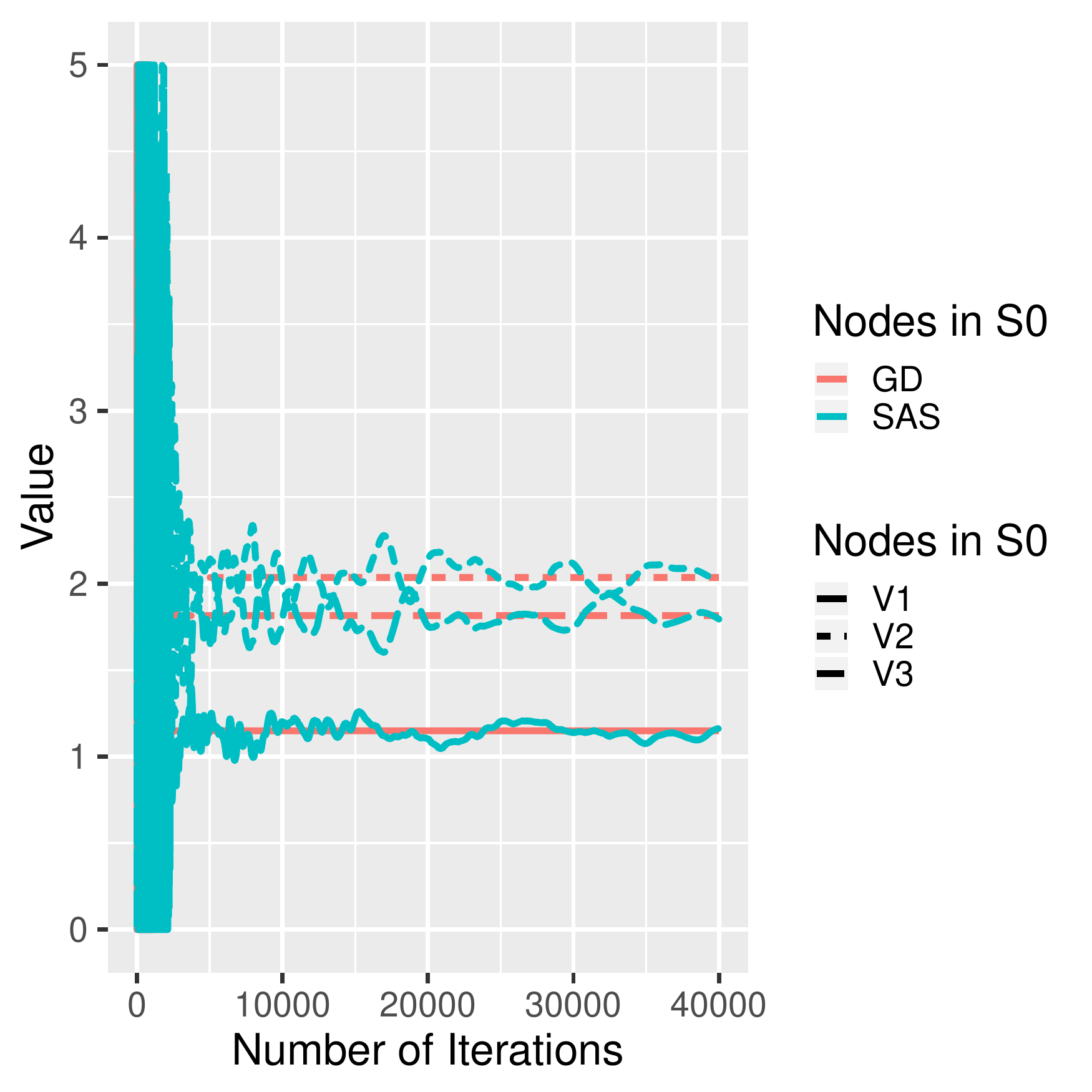}\label{fig:annel1000}}
 \subfigure[$denom=5000$.]{
\includegraphics[scale=0.3]{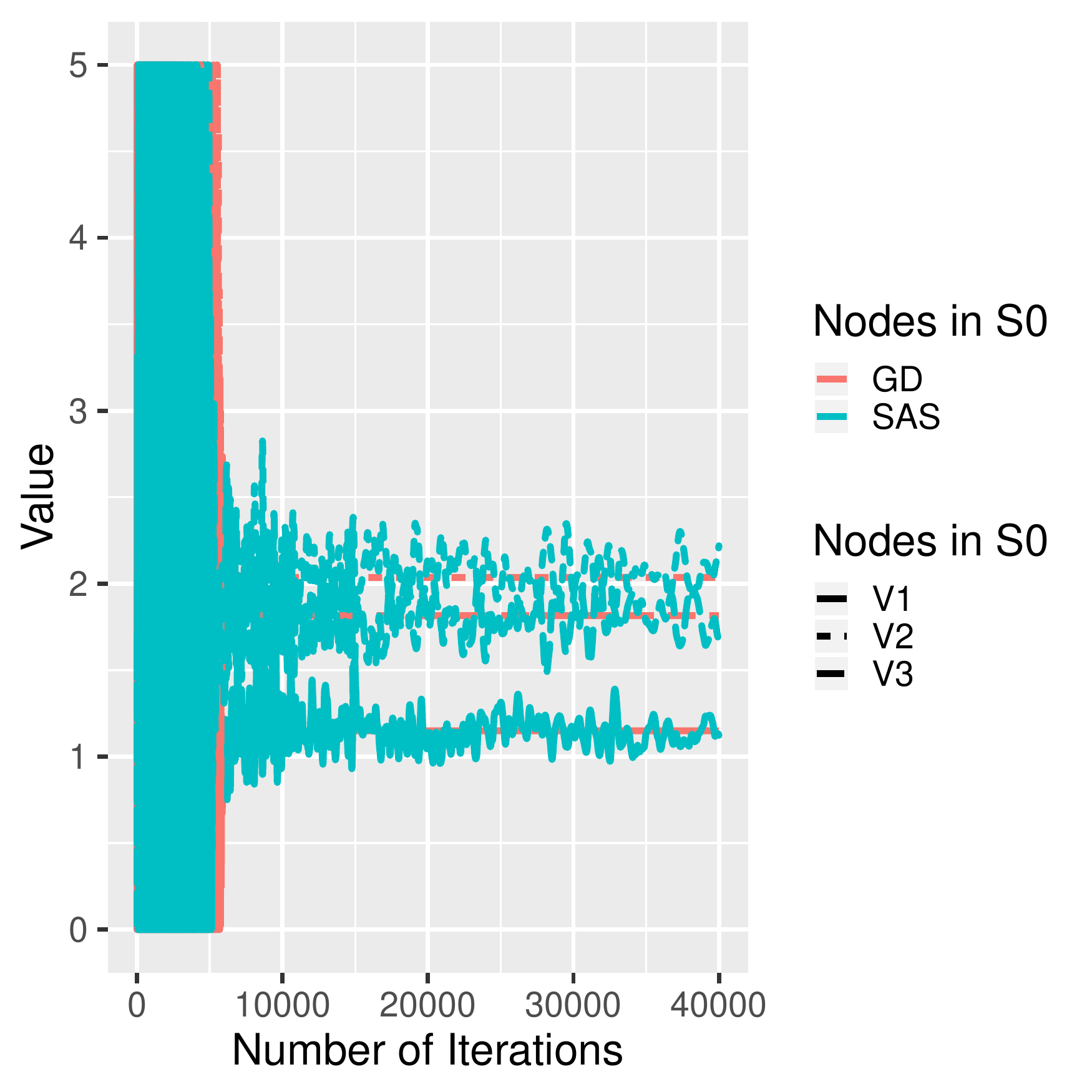}\label{fig:annel5000}}
\caption{Convergence of the annealing scheme with/without the reinforcement learning scheme.}
\end{figure}

\noindent\textbf{Study of the more general problem:} The final numerical study is dedicated to the last reinforcement scheme based on annealing method for non-convex optimization. We restrict this study to the Karate network.
We assume that $w_i(u_i)=h(i)$ for all $i\in\SA$ and $u_i\in[0,M]$. Also in this simulation study for $i\in S$, $\alpha_i(u_i)=\frac{u_i}{u_i+0.1}$ with $\#S=3$. The noisy term of \eqref{eq: dec gm 2} is parametrized by $c(k):=\lceil k/denom \rceil$ and $C=10$.
We study two schemes. The first one is reinforcement learning  (\eqref{eq: approx grad gm 1}, \eqref{eq: approx grad gm 2} and \eqref{eq: dec gm 2}). The second one is without the approximation of the reinforcement scheme for the computation of the gradient (\eqref{eq: approx grad gm 3} and \eqref{eq: dec gm 2}).
 We are interested in understanding how the first scheme tracks the behavior of the second scheme. In Figure \ref{fig:annel1000} (resp. Figure \ref{fig:annel5000}) , we observe the first algorithm starts to track the trajectory of the second trajectory after 2000 iterations (15000 iterations). Moreover, we observe that in both cases, the two schemes are converging to the same values.

\section{further directions}\label{sec: further directions}
 \noindent\textit{Incorporating subjective risk measures:} Since we are modeling social networks, it is desirable that we incorporate behavioral aspects into our model explicitly, such as the risk-measures suggested by behavioral economics. This makes the problem a lot harder, see, e.g., \cite{Shen} for some initial efforts towards the dynamic programming aspects.

\noindent \textit{How to select the initial set of agents:} One of the results of this paper is the fact that by observing a small number of agents, we can increase drastically the speed of convergence of the algorithm. Even if the obtained solution is suboptimal,  the relative difference observed between the optimal payoff and the suboptimal one, in the simulation, was low (about 0.01\%). Therefore, one  interesting question  would be to find a possible algorithm to choose the initial set of agents? This question can be related to the problem of selecting sensors $k$, among $n$  potential sensors. In future work, we will try to adapt this well-known problem to our setting. See also a greedy scheme for agent selection with performance guarantees proposed in \cite{Consent}.

 \noindent\textit{Other learning schemes:} In our current scheme, we observe communications between a set of particular agents. In \cite{banerjee2017using}, the authors prove that agents in a social network can easily guess who is central in a diffusion process. Therefore a potential scheme would be to ask a small number of agents who they think is central in the network and factor this information into the opinion-shaping optimization problem.

\noindent \textit{Pricing scheme for accessing communication data:} Accessing the data in the age of information is getting more and more important. Agents start to realize the value of their data.  The question that we should ask in our setting is the following: how much should a planner pay an agent to access her information in order to be able to perform opinion shaping, i.e., design an incentive-compatible pricing mechanism for data acquisition.

\section*{Acknowledgment} The work of VSB was supported in part by a J.\ C.\ Bose Fellowship from the Department of Science and Technology, Government of India, and the project `\textit{Machine Learning for Network Analytics}' from the joint DST-INRIA program administered by the Indo-French Centre for Promotion of Advanced Research.

\bibliographystyle{plain}
\bibliography{ref}

\end{document}